\documentclass[12pt]{article}
\usepackage{epsfig}
\begin{document}

\title{\bf  Fractal dust model of the Universe based on
 Mandelbrot's Conditional 
Cosmological Principle and General Theory of Relativity}

\author{A. K. Mittal\thanks{\small Harish Chandra
 Research Institute, Jhusi,
Allahabad; and Department
 of Physics, 
University of Allahabad, Allahabad - 211 002, India; 
mittal\_a\_k@indiatimes.com; mittal@mri.ernet.in;}, 
Daksh Lohiya\thanks{\small 
Department of Physics and Astrophysics, University of
 Delhi, 
New Delhi--7, India; dlohiya@ducos.ernet.in}}

\date{}
\maketitle 

\vspace{-1.2cm}
\begin{center}
\em Inter University Centre for  Astronomy and Astrophysics,
Postbag 4, Ganeshkhind, Pune 411 007, India
\end{center}

\begin{abstract}

	We present  a fractal dust model of
 the Universe based on
 Mandelbrot's proposal to replace the standard
 Cosmological 
Principle by his Conditional Cosmological Principle within the framework
 of General Theory of Relativity. This
 model turns out to be free from the de-Vaucouleurs
paradox and is consistent with the SNe1a observations. The expected
galaxy count as
a function of red-shift is obtained for this model. An interesting
variation is a steady state version, which can account for an
accelerating scale factor without any cosmological constant in the model.

\end{abstract}

\vspace{.5cm}
{\bf Running Title: Fractal dust model of the Universe}

\pagebreak
\section*{} 
\vspace{.5cm} 
{\bf1.~~INTRODUCTION}
\vspace{.5cm}
 
	Fractality is ubiquitous in nature. Should Cosmology
be an exception? Mandelbrot's vision, followed by elaborate
demonstration of fractality in galaxy distributions by
Pietronero, have perhaps been singularly responsible for
ensuring that any modern text on cosmology is rather 
incomplete without at least a chapter on Fractals. Unfortunately, 
it stops here as there is no definite ansatz that could
be used to match cosmological predictions. This article explores a 
way out of this impasse.

	Standard cosmology is based on the assumption of
 homogeneity 
 and isotropy of the Universe, the so-called Cosmological
 Principle, on
  scales greater than $\approx 10^8$ light years. The
 Friedmann metric, and 
consequently the Hubble Law,
 follow from this assumption. Although  the metric is
 expected to be 
valid only on scales larger
 than the  scale of homogenization, the  Hubble law is
 found to hold on
 smaller scales. This seeming contradiction goes by the
 name of the
 `Hubble-deVaucouleurs' [H-deV] paradox$^1$. On
 the other hand, during the
 last decade, the assumption of homogeneity
on scales greater than $10^8$ light years itself has come
 to be challenged$^{1,2}$. 
It is now believed 
that the scale of homogenization, if any, is definitely
 greater than 100 Mpc. 
The number of galaxies $N(r)$ within a sphere of radius
 $r$ centred on {\it any} galaxy, is not
proportional to $r^3$ as would be expected of a
 homogeneous distribution. Instead $N(r)$ is 
found to be proportional to $r^D$, where $D$ is
 approximately equal to 2. Without assigning
a special central position to an observer, such a scaling
 can be explained only by assuming
that galaxies are distributed on points belonging to a
 fractal set of dimension D.  
  	It has further been argued that available evidence
indicates that fractal distribution of visible matter
 extends
well upto the present observational limits without any
 evidence of 
cross-over to homogeneity$^{1,3}$
 This
 suggests that the entire Universe could
be a fractal. At present this question is being hotly
 debated$^{1-6}$ 
  	
  	No generally acceptable structure formation
 scenario, that could explain
the observed inhomogeneities, has yet emerged, either
 within
the framework of the standard big-bang cosmologies
$^{7,8}$
or alternative cosmologies such as the quasi-steady state
 cosmology$^9$
  	
	 Although, a fair amount of evidence has been
  collected in support of inhomogenous distribution of
 visible matter, there 
  is no evidence to contradict isotropy in a statistical
 sense from any
  galaxy. In view of the observed fractality and 
isotropy,  Mandelbrot$^{10}$
 proposed the replacement of the standard Cosmological
  Principle by the Conditional Cosmological Principle.
 According to this
  principle the Universe appears to be the same
 statistically from every 
  galaxy (point of the fractal) and in every direction.

	 In Sec 2, we
  present a model fractal Universe that is based on the
 Mandelbrot's Conditional
 Cosmological Principle and the General Theory of
 Relativity. We show how the Conditional Cosmological 
Principle leads to the Friedmann metric and how the
 Einstein equation can be satisfied in the fractal context.

	In Sec 3, we obtain the time dependence of the
scale function for two cases. If we assume that the number
 of galaxies is
conserved, we obtain the FRW metric with a non-zero
 effective density, whereas 
the average density for the fractal Universe is zero. Thus
the redshift-distance relation in the fractal model turns
 out to be the same
 everywhere as that over scales 
greater than the homogenization scale  in the
 standard model. Hence this fractal model is free from the
 Hubble-deVaucouleurs paradox. On the other hand, if we
 assume that the large scale
 fractal distribution of galaxies
is in a steady state, we obtain  an accelerating Universe
 whose acceleration
is related to the fractal density and the Hubble's
 constant.

 It has previously been
argued $^{11}$ that fractal scaling can be 
obtained along
 the past light cone in a perturbed Einstein -
de Sitter Cosmology so any observed fractality is not
 necessarily inconsistent with
the Cosmological Principle and there is no
 Hubble-deVaucouleurs paradox. In Sec 4, we obtain
galaxy counts along the past light cone on the basis of
 our model, assuming that the `effective density'/
`fractal density' can be neglected. Fractal
scaling on a constant time hypersurface implies a fractal
 scaling 
along the past light cone for small
red-shifts.

	It is pertinent to recall that the asociation of
 the FRW metric obtained
from smoothed out homogeneous Universe and the actual
 Universe is not clearly
established to date. This goes to the root of ``the
 averaging problem'' 
 in General Theory of Relativity
 $^{12-14}$.
Once the manner in which the FRW metric holds is
 established, it may have an 
essential bearing on the inferences drawn from the
 fractal model in this paper.

	In Sec 5, we discuss the previous attempts at
reconciling the observed fractal structure with a 
relativistic description of the Universe and how the
 Conditional Cosmological Principle, as used in our model,
leads to a more satisfactory picture.

	Conclusion is presented in Sec 6.

\vspace{.5cm}    

{\bf 2.~~THE FRACTAL MODEL}
\vspace{.5cm} 

 	The Cosmological Principle provides the symmetry
 necessary to derive the FRW metric.
Mandelbrot's Conditional Cosmological Principle weakens
 the Cosmological Principle as
it demands that the Universe appears statistically the
 same to all observers situated on 
a galaxy (point of a fractal) but not in a region of
 void. More specifically,$^{15}$ 
in a reference frame with origin P,
the distribution of matter is independent of P under the
 sole condition that P must be a 
material point. If P is not a material point and R is
 fixed, a sphere of radius 
R centred on P is empty with probability equal to one.
 
 	 Just as the standard model follows naturally
 from the
Cosmological Principle when General Theory of Relativity
 is applied, an ansatz for a fractal
model follows naturally from Mandelbrot's Conditional
 Cosmological Principle, once the
necessary change in perspective required to deal with
 fractal distributions is made.
This ansatz is based on the observation that in a fractal
 universe, density is not 
defined at any point.
 Hence Einstein's equations do not mean anything at a
 point. However, by replacing 
the concept of density at a point by that of a conditional `mass
 measure'
defined over sets, it is possible to formally satisfy the
 Einstein's equations integrated over sets.
Conditional Cosmological Principle then means that the conditional
mass measure will be the same for all observers situated at points
belonging to the fractal.
    	 
 	We define a ``hypersurface of homogeneous
 fractality of
  dimension $D$'' as the hypersurface in which the mass
 measure over a
  sphere of radius $R$ centred on the observer is
 proportional to $R^D$.
  We say that the Universe is a fractal universe of
 dimension $D$, if
  through each galaxy in the Universe, there passes a
 spacelike
  ``hypersurface of homogeneous fractality of dimension
 $D$''. 
  
  	Isotropy of the Universe means that, at any
 event, an observer
  who is at rest in this hypersurface cannot
 statistically distinguish any
  space direction from another. 
  
  	It is widely believed that isotropy from all
 points of observation
  implies homogeneity. Thus an inhomogeneous Universe
 like a fractal could
  not be isotropic. It was shown$^4$ from
 the observed 
isotropy of the Universe
  that the fractal dimension of the Universe could not
 differ appreciably from
  $3; ~~(|D-3|< 0.001)$. However, 
  Mandelbrot $^{10,15}$ has demonstrated how 
  to construct fractals of any given dimension whose
 lacunarity
  could be tuned at will to make the distribution as
 close to isotropy
  (from any occupied point of the fractal) as desired.
Thus in a fractal
  scenario isotropy from all galaxies does not rule out
 surfaces of
  homogeneous fractality. 
  
  	Isotropy of a fractal universe implies that the
 world lines
  of the cosmological fluid are orthogonal to each 
hypersurface of 
  homogeneous fractality. This allows the slicing of 
spacetime into hypersurfaces 
of constant time as for the standard model.

	For the standard homogeneous model the
 Cosmological Principle leads
to the Friedmann metric $g_{\mu\nu}^{FRW}$ specified by
 the line element:
$$
ds^2 = g_{\mu\nu}^{FRW} dx^\mu dx^\nu  = -dt^2 + a^2(t)\{ d\chi^2 + \Sigma^2(\chi)(d\theta^2 + sin^2\theta d\phi^2)\}\eqno{(1)}
$$
where
$$
\Sigma(\chi) =\left\{ \begin{array}{lr} \sin (\chi) & for~positive~spatial~curvature,~k=1 \\ 
\chi & for~zero~spatial~curvature~ k=0 \\ 
\sinh(\chi) & for~negative~ spatial~ curvature,~ k=-1\end{array} \right.\eqno{(2)}
$$
This metric yields the component $G^{00}$ of the
 Einstein tensor,
$$
G_{FRW}^{00} = 3\{({{\dot{a}}\over {a}})^2 + {k\over {a^2}}\}\eqno{(3)}
$$
As demanded by the Cosmological Principle, this has the
 same value everywhere
on a hypersurface of constant time.

	For a fractal Universe, the Conditional
 Cosmological Principle demands that
on any given hypersurface $S$  of constant time, $G^{00}$ 
has the same value at the points
of the fractal and zero elsewhere. This is consistent
 with the Einstein Equation
$$
G^{00} = 8 \pi \rho\eqno{(4)}
$$
because for a fractal
$$
\rho (P)  =  \Sigma_i m \delta(P,P_i).\eqno{(5)}
$$
Here $P_i$ denote the points of the fractal each having
 mass $m$. $G^{00}$ denotes the
moment of rotation density, which is the sum of intrinsic
 and extrinsic curvature at the point P $^{20}$.
With every point of mass $m$ is associated a moment of
 rotation $ 8\pi m $.

	If $S_P^3 (R)$ denotes a hypersphere of radius
 $R$ centred at a point $P$
belonging to a fractal of dimension $D$, then
$$
\int_{S_P^3(R)} \rho dV = \int_{S_P^3(R)} d\mu = M_P(R) = C(t)R^D\eqno{(6)}
$$
	The discrete mass measure $d\mu = \rho dV$ may be
 replaced by a smoothed out conditional 
measure ${\hat \rho} dV = (D/4\pi)C(t)r^{D-3} dV$.

	As is clear from the above, for a fractal
 distribution of matter, the concept of
density is undefined and has to be replaced by the notion
 of a measure on sets.
This implies that $G^{00}$, the moment of rotation
 density,
 is not defined at any point.
However, over any set in  constant-time surface, we must
 have
$$
\int G^{00} dV = \int d\mu_{MR} = 8\pi\int d\mu\eqno{(7)}
$$

 The exact moment of rotation measure $d\mu_{MR}$ may be replaced
by the smoothed out conditional measure 
$\hat G^{oo}_{fractal} dV$, for an observer at point $P$, where, 
$$
\hat G^{oo}_{fractal}(t,\chi,\theta,\varphi) = \left\{\begin{array}{lr}
\hat f(\chi ) G^{00}_{FRW}(t)& if~ P~\in~  the fractal\\
0 & otherwise\end{array}\right. \eqno{(8)}
$$
Then,
$$
4\pi G^{00}_{FRW}(t)a^3(t)\int^\chi_0 \hat f(\chi)\Sigma^2(\chi) d\chi = 
8\pi C(t)a^D(t)\chi^D \eqno{(9)}
$$
This is satisfied by		
$$
G^{00}_{FRW}(t) = 2\nu C(t)a^{D-3}(t)\eqno{(10)}
$$
$$
\hat f(\chi) = {D\chi^{D-1}\over {\nu\Sigma^2(\chi)}}\eqno{(11)}
$$
The value of $\nu$ is determined by demanding that $\hat f(\chi)$ = 1 for 
$k = 0$ and $D = 3$. This gives $\nu$ = 3. 
$\hat G^{~00}_{fractal}$ satisfies the 
integrated Einstein equation over a sphere of radius $R$. 
Here $\hat G^{~00}_{fractal}$ is not a function. It is
 an ansatz for 
defining a smoothed out moment of rotation measure
on sets containing the point $P$ just as the mass
 measure is 
expressed by using the smoothed out measure $\hat\rho$
 proportional to $r^{D-3}$. Here $\hat\rho$
 does not mean the density at a point, but
 merely an ansatz to compute the mass measure. In this
 way 
Einstein's equations for a fractal distribution of mass
 are expressed by a relation
connecting $\hat G^{00}_{fractal}$ to $\hat\rho$,
 remembering clearly that these are not
functions but ansatz to compute conditional measures.
 Thus the dependence of
 $\hat G^{00}_{fractal}$ on $\chi$ and of $\hat\rho$ on
 $r$ should not be seen as an 
indication of inhomogeneity but rather as a means of
 concrete realization of the 
 Conditional Cosmological Principle.

	 The averaging procedure over a constant time
 hypersurface used here, 
is in fact tacitly assumed in the standard model while
 making the fluid approximation. 
Both for the homogeneous model and the fractal model,
 the dust mass distribution is
the sum of delta functions. In one case, these delta
 functions are distributed homogeneously
whereas in the other they are distributed on a fractal
 set.
In both cases Einstein's equations can be satisfied only
 when integrated over sets as they
are otherwise ill defined for point mass distributions.
 This integration essentially sums over
discrete masses for the `matter' side of the Einstein's 
equations and over discrete moments of rotation
 for the `geometry' side of the Einstein's equations. In
 both cases, 
a clumpy matter distribution and a clumpy geometry are
 smoothed out in a manner that integrations 
over finite sets give the same result as for the clumpy
 case. Cosmological principle in the
case of homogenous distributions and Conditional 
Cosmological Principle in the case of fractal
 distribution allow simplified descriptions of the
 Universe. 

	The above argument can be made more explicit by 
assuming that the Universe 
is made up of homogeneous
galaxies of mass $M_g$ and radius $R_g$. First, let us 
consider the case of a homogeneous distribution 
of these galaxies. It is clear from the Einstein eqn
 $G^{00} = 8\pi\rho$ that
$$
G^{00}(P) = \left\{\begin{array}{lr}
{3\times 8\pi M_g}\over {4\pi{R_g}^3} & if ~~ P ~\in~ some~galaxy\\
 0 & otherwise\end{array}\right.\eqno{(12)}
$$ 
Let us call this function $G^{00}_{exact}$.
Suppose there are $N_{hom}$ galaxies in a sphere of
 radius $R$. Then,
$$
	\int_{S^3(R)} G^{00}_{exact} dV = 8 \pi N_{hom} M_g\eqno{(13)}
$$
	In the fluid approximation, the discrete mass
 distribution is replaced by a smoothed density
distribution. For this distribution,
$$
	\int_{S^3(R)} G^{00}_{smooth} dV = 8 \pi N_{hom} M_g  \eqno{(14)}
$$
	It is clear that $G^{00}_{exact}$ is not equal
 to $G^{00}_{smooth}$. The metric coefficients
$g_{\mu\nu}^{exact}$ that would give rise to
 $G^{00}_{exact}$ will be different from the metric
coefficients $g_{\mu\nu}^{smooth}$ which give rise to
 $G^{00}_{smooth}$. The FRW metric gives the
$g_{\mu\nu}^{smooth}$ and inferences about red-shift etc
 are drawn from it. It is assumed
that these inferences hold for the exact distribution.

	For our model fractal Universe, suppose there
 are $N_{frac}$ galaxies in a sphere of radius
$R$. Then
$$
\int_{S^3(R)}\hat G^{00}_{frac-exact} dV = 8 \pi N_{frac} M_g = \int_{S^3(R)}\rho dV \eqno{(15)}
$$
We see that  $G^{00}_{frac-smooth} = f(\chi) G^{00}_{FRW}$ satisfies
$$
\int_{S^3(R)} G^{00}_{frac-smooth} dV = 8 \pi N_{frac} M_g =  \int_{S^3(R)}\rho dV \eqno{(16)}
$$

	To deal with fractal distributions the Einstein
 equation may be generalized to
$$
  \int_{S^3(R)} d\mu_{MR} =  8\pi \int_{S^3(R)} d\mu\eqno{(17)}	
$$
where $d\mu_{MR}$ denotes a measure for moment of rotation.
 For a homogeneous distribution
 $d\mu_{MR} = G^{00}_{FRW} dV$ and $d\mu = \rho dV$. For
 a fractal distribution
 $d\mu_{MR} = \hat f(\chi) G^{00}_{FRW} dV$ and $d\mu = \hat \rho dV$.

\vspace{.5cm} 

{\bf 3.~TIME DEPENDENCE OF THE SCALE FACTOR}

\vspace{.5cm}

	We obtain the time dependence of the scale factor
in two cases:
\vspace{.5cm} 

{\bf 3.1 Case 1: Conserved galaxy number:} 
\vspace{.5cm} 
   
	In this case we assume that 
the number of galaxies remains unchanged as the scale
 factor changes with time. Then $C(t)a^{D}(t){\chi}^{D} = C(t_0)a^{D}(t_0){\chi}^{D}$ 
so that
$$
C(t) = {a^D(t_0)\over {a^D(t)}} C(t_0)\eqno{(18)}
$$

	In this way, we get,
$$
3\{({{\dot a}\over a})^2 + {k\over {a^2}}\} = 6{{a_0^D}\over {a^3}}C_{a_0}
\eqno{(19)}
$$
where $a_0$ is the scale factor at time $t_0$.

	The dynamics of the scale factor due to a
 fractal distribution of matter, 
which satisfies 
the conditional cosmological principle is the same as in
 standard cosmology for
homogeneously distributed matter with an effective density
$$
\rho_{eff}  =  {3\over {4\pi}} {{a_0}^{D}\over {a^3}}C_{a_0}\propto {1\over{a^3}}\eqno{(20)}
$$
It should be noted that although the time dependence of
 the scale factor is the
same as for a homogeneous Universe, the effective
 density for the fractal Universe is
different from the average density which is zero.

	The value of the `fractal density' $C_{a_0}$ at
 the present epoch $t_0$, may be obtained
from the observed number of galaxies in a sphere of
 radius $R$. Then the scale factor for
the present epoch may be obtained from 
$$
		{H_0}^2  + {k\over{{a_0}^2}} = 2 {a_0}^{D-3} C_{a_0}\eqno{(21)}
$$
	
From the values $C_{a_0}$ and $a_0$ the proportionality 
constant in eqn(20) 
can be determined. For $D = 2$,
$$
a_0 = {{C_{a_0} + \sqrt{C_{a_0}^2 -kH_0^2}}\over {H_0^2}}\eqno{(22)}
$$
   
	From the galaxy number count data 
of Labini et al$^1$, the average conditional
 number density ${\Gamma}^{*}$ of
galaxies over a radius of 100 Mpc is $\approx 10^{-2}(Mpc)^{-3} $. The total number 
of galaxies in a sphere of radius R is given by:
$$
N(R) = nR^2\eqno{(23)}
$$
where $n$ is the ``fractal number density''. This gives
the average conditional number density
$$
{\Gamma}^{*} = {{3N}\over {4\pi R^3}} = {{3n}\over {4\pi R}}\eqno{(24)}
$$
One therefore gets $n\approx 4(Mpc)^{-2}$. Taking a
 typical galaxy mass as 
$\approx 1.8\times 10^{11} M_\odot$, gives $C_{a_0} \approx 10^{-4}~ gms~cm^{-2}$. In 
gravitational units this amounts to $C_{a_0} \approx 2\times 10^{-24}~sec^{-1}$. This is
small in comparison to the observed Hubble parameter
 $H_0 \approx 2\times 10^{-18}~sec^{-1}$.   
Such a universe would be curvature dominated even for
 redshifts as high as $10^5$ and
its coasting would be indistinguishable from a linear
 coasting Milne model: $a(t) = t$
at lower redshifts.

	It is straightforward to put this scaling to
 classical cosmological tests, viz.:
(1) The galaxy number count as a function of redshift;
 (2) The angular diameter
of ``standard'' objects (galaxies) as a function of
 redshift; and finally
(3) The apparent luminosity of a ``standard candle'' as
 a function of
redshift. The first two tests are marred by evolutionary
 effects and for
this reason have  fallen into disfavour as reliable
 indicators of 
a viable model. However, the discovery of Supernovae
 type Ia [SNe Ia] 
as reliable 
standard candles  has raised hopes of elevating 
the status of the third test to that
of a precision measurement that could determine the
 viability 
of a cosmological model. The main reason for regarding
 these objects
as reliable standard candles are their large luminosity,
 small dispersion 
in their peak luminosity and a fairly accurate modeling
 of their 
evolutionary features.

For a linearly coasting model, the apparent 
magnitude of an object is related
 to its  redshift $z$ by:
$$
m =  25 + M  + 5log[a_o Sinh(\chi) (1 + z)]\eqno{(25)}
$$
It is straightforward to reduce it to
$$
  m(z) = 5 log(\frac{z^2}{2} + z) + \mathcal{M} \eqno{(26)}
$$
with ${\mathcal M} \equiv M -5log(H_0) + 25$

	Figure `1'$^{17,18}$
sums up  the  Supernova Cosmology project data 
for supernovae with redshifts
between 0.18 and 0.83 together with the low redshift set 
at redshifts below 0.1. Also plotted is the latest 
SNe1a at redshift 1.7 [see eg. Wright$^{18}$]. 
Clearly, the Fractal model described here is as good 
a fit as the constrained Standard Cosmology model with $(\Omega_\Lambda, \Omega_M) 
= (.0.72, 0.28)$. The goodness of concordance can be
 judged by the fact that the $\chi^2$ 
per degree of freedom is roughly unity for the fit.
 As a matter of fact a linear coasting is accommodated
 even in
the 68\% confidence region. This finds a passing mention
 in the analysis of
Perlmutter$^{17}$
who noted that the curve for $\Omega_\Lambda = \Omega_M = 0$ (for which
the scale factor would have a linear evolution), is 
``practically identical to the best fit plot for an
 unconstrained cosmology''.

	It is interesting to note that, unlike the standard model, for 
any observed set of values $H_0$ for the Hubble constant and $C_{a_0}$
for the fractal density, eqns (19) and (21) also admit a $k = 0$ solution
for appropriate choice of $a_0$.

\vspace{.5cm} 
{\bf 3.2: Case 2: Steady State Fractality}:
\vspace{.5cm} 

	 Another interesting model may be obtained if we
 assume that the number of galaxies in
a sphere of radius $R$ is the same at all epochs so that
 $C(t)$ is a constant. This
gives a fractal version of the steady state model,
 although it lacks the maximal symmetry
in space-time of the ``Perfect Cosmological
 Principle''.

	In this case we get,
$$
 		3\{({{\dot a}\over a})^2 + {k\over {a^2}}\} = 6 C a^{D-3}\eqno{(27)}
$$
For $D = 2$, we obtain,
$$
	({{\dot a}\over a})^2 + {k\over {a^2}} = 2 {C\over {a}}\eqno{(28)}
$$
The solution of this equation is
$$
	a(t) = a_0  +  H_0 a_0 (t - t_0) + {C\over 2} (t - t_0)^2\eqno{(29)}
$$
where $a_0$ is the scale factor at the present epoch $t_0$ and satisfies the equation,
$$
{H_0}^2 {a_0}^2 + k =  2 C a_0\eqno{(30)}
$$
For $k = 0$ the deceleration parameter is given by
$$
q = - {{\ddot{a_0} a_0}\over{{\dot a}^2}} = -{C\over {a_0H_0^2}} = -{1\over 2}\eqno{(31)}
$$
For $k = -1$,
$$
q = - {C\over{C + \sqrt{C^2 + H_0^2}}} \eqno{(32)}
$$
If $C << H_0,~~ q$ is approximately equal to $-C/H_0$. 
This is a rather low value and its
concordance again coincides with that of the 
empty model.

	However, it should be noted that the Conditional
Cosmological Principle could be the consequence of an 
underlying fractal structure of space-time, in which case any
dark matter wold also have the same fractal structure as visible
matter. This steady-state model therefore offers the possibility
of accommodating the acceleration of the scale factor without
invoking any cosmological constant.
  
\vspace{.5cm}
 
{\bf 4. RED-SHIFT DEPENDENCE OF GALAXY COUNTS} 
\vspace{.5cm} 

	As astronomical observations take place on the
 past light cone, the number count 
of galaxies inside a hypersphere of radius
 $ R = a(t)\chi$ on a constant time hypersurface
is unobservable. The red-shift dependence of galaxy
 counts can be derived as follows:

	We assume that $C_{a_0} << H_0$ so that,
$$
({\dot{a}\over {a}})^2 - {1\over {a^2}}  =  0\eqno{(33)}
$$
For the present epoch this gives  $a_0 = H_0^{-1}$ so that 
the time dependence
 of the scale factor is given by
$$
a(t) = (t - t_0) + {1\over {H_0}}\eqno{(34)}
$$  
	If a light ray is emitted from a source at event
 $(t_e,\chi_e ,\theta , \phi)$ and received
by an observer at event $(t_0, 0, \theta ,\phi)$, then
 along the light cone, we must have
$$
{{dt}\over {d\chi}} = - a(t) = -(t - t_0 + {1\over{H_0}})\eqno{(35)}
$$
so that
$$
\chi = - ln \{H_0(t - t_0) + 1\}\eqno{(36)}
$$
The red-shift is given by
$$
1 + z  =  {{a(t_0)}\over{a(t_e)}} = e^{{\chi}_e}\eqno{(37)}
$$
Number of galaxies having $ \chi < {\chi}_e $ is given by 
$ C_0 {a_0}^D {{\chi}_e}^D$.  
Therefore the number of galaxies having red-shift less
 than z is given by:
$$
N(< z) = {C_0\over {{H_0}^D}}[ln(1+z)]^D\eqno{(38)}
$$

For small z, fractal scaling is seen along the
 past-light cone. For larger z 
the deviations from pure power law behaviour may be
 compared with observational data to test the model.

\vspace{.5cm} 
{\bf 5. DISCUSSION}
\vspace{.5cm} 

There have been several attempts to incorporate
 large-scale inhomogeneities of the 
Universe in the framework of General Theory of
 Relativity. However, they have not exploited
the Conditional Cosmological Principle proposed by
 Mandelbrot as a replacement of the 
 Standard Cosmological Principle.

In case future observations unambiguously demonstrate
 that the Cosmological
Principle is not valid, entire standard cosmology
 scenario will break down. In such an
eventuality our model based on Mandelbrot's Conditional
 Principle provides 
the simplest alternative around which modified 
cosmologies may be built. 
For the present, this 
model can help resolve many of the vexing questions of 
relevance to the `fractal debate'.

 	Abdella et al$^{11}$  have suggested
that fractal scaling observed by Pietronero
 and coworkers 
$^{1,2}$ is
simply an apparent scaling due to the fact that
 observational quantities such
as density lie along the past light cone and depend more
 significantly on the
red shift than had hitherto been assumed. The average
 density along the past light
cone becomes inhomogeneous, even in the spatially
 homogeneous spacetime
of standard cosmology. However, it does not have the
 observed fractal scaling.
By introducing perturbations, they could obtain an
 approximate scaling. In 
this way they tried to reconcile the observed fractal
 scaling with the
standard Csomological Principle.

	Compared to this procedure, our model based on
 the Mandelbrot's
Conditional Cosmological principle derives the galaxy
 count scaling law 
along the past light cone in a simple straight forward
 manner. This scaling
agrees with the observed fractal scaling for low
 red-shifts. If Conditional
Cosmological Principle is to hold along the same lines 
as the standard
Cosmological Principle, the fractal scaling has to hold 
along the constant
time hypersurface and not along the past light cone. For 
low red-shifts there
is negligible difference between the two. Deviations at 
higher red-shifts
may be used to test our model.

	From the apparent fractal conjecture
 perspective, the
Hubble-deVaucouleurs paradox is resolved by attributing 
apparent
fractality to observations of a perturbed FRW universe 
on the
past light cone. It is claimed that there is no 
inconsistency between 
apparent fractal scaling and the observed linear Hubble 
Law on scales smaller than the homogenization scale.

	In the approach of this paper, the Conditional 
Cosmological
Principle forces the points of the fractal to follow the 
Hubble 
flow, that is, to remain at rest in the comoving 
coordinates. Thus
the expected increase in peculiar velocities with greater 
inhomogeneities observed on larger scales is not to be 
found. 
Hitherto all attempts
to treat the fractal structure in a relativistic context
have explicitly or implicitly regarded the fractal 
structure
as inhomogeneities, with a background homogeneity 
providing
the relativistic framework in the form of Friedmann 
metric. 
In our approach, the homogeneous background is replaced by
 homogeneous fractality. In the fractal picture, there 
is no
average density and therefore no inhomogeneity. All the
points of the fractal are on equal footing, each at rest 
in
the comoving coordinates. The observed linear
Hubble Law is consistent with the observed fractal scaling
thus resolving the  Hubble de-Vaucouleurs paradox.

	The apparent fractal conjecture is based upon an
inhomogeneous spherically symmetric metric. This does not
put all the points on equal footing and the derived 
scaling
would hold only from one point, the centre of spherical 
symmetry.
Contrary to this, the observed fractal scaling of galaxy
 distribution, is 
a power law scaling from every galaxy. Only non-analytic
fractal sets can give rise to this kind of scaling. It is
not sufficient to obtain power law scaling from one 
point to
claim that the observed fractal scaling has been explained
without giving up the standard Cosmological Principle. 

	Being based on the Mandelbrot's Conditional 
Cosmological
Principle, the model presented here puts all the  points 
of the
fractal on equal footing. It deals with
non-analytic distribution of matter in the General Theory
of Relativity framework with the help of conditional 
measures.
 Non-analytic distribution of matter will necessarily
be associated with a non-anlytic space-time geometry. 
Till a
totally satisfactory mathematical framework for dealing 
with
fractals emerges, one has to try to deal with them using
smoothing methods. However, in the case of fractals the 
smoothing
has to be identical from every point of the fractal. The 
use
of smoothed out conditional measures is our suggestion 
to achieve this.

	Further, the basic assumption of the apparent 
fractal conjecture is
that the fractal scaling law deduced by Pietronero and
 coworkers$^{1,2}$ is
based on taking the Euclidean space approximation, so
 that no 
distinction has
been made between the observable past light cone and the 
unobservable 
constant time
hypersurface. This does not appear to be
correct. Labini et al $^{1}$ have clearly stated 
that comoving distances have been 
computed by using the Mattig formula for q = 1/2. It has 
also been stated
that the use of different values of q does not have 
significant effect on
the results for small red-shifts. The justification of 
this procedure, in the absence of a relativistic
 framework for fractal cosmology, is another matter. 
For the Conditional
Cosmological Principle and our model to apply, it is 
necessary that fractal
scaling on a constant time hypersurface exists upto very 
large scales, so that the
fractal can be treated as infinite. Comparison of the 
red-shift 
dependence of galaxy counts derived
for our model with the observed data for 
moderate and large red-shifts, would provide another 
test for the model.

	From the above arguments, we feel that the model 
of this paper,
which may be called the {\it homogeneous fractal expanding model}, 
is better suited than
the apparent fractal conjecture to provide a starting 
point for 
developing a theoretical
framework that can replace the standard framework.      
    
\vspace{.5cm} 
{\bf 6. CONCLUSION}
\vspace{.5cm}
 
	The Standard Cosmological Principle is not 
merely an esthetically pleasing and 
philosophically satisfying principle; it plays a crucial 
role in developing the frame
work of Standard Cosmology. However, Standard Cosmology 
has not been able to satisfactoriy 
explain the observed large scale distribution of 
galaxies, which seems to satisfy a fractal 
scaling law upto the largest scales investigated.

	 On the other hand the fractal scenario
till now had no satisfactory explanation even for 
observed red-shift of galaxies.
It has generally been believed that if the Universe 
would be hierarchical then there is no
known analysis of redshift data that is self-consistent 
and if the Cosmological Principle
could be shown to be false, then cosmology would not be 
the coherent body of knowledge
that many theorists believe that it 
is$^{19}$. However, the use of Mandelbrot's 
Conditional
 Cosmological Principle in the framework of
General Theory of Relativity, as described in the model 
presented here,
 provides the means to explain the observed red-shifts 
of galaxies. If the Cosmological
Principle is eventually shown to be false, the 
Conditional Cosmological Principle may
provide cosmology with a theoretical underpinning 
necessary for the analysis and 
interpretation of observational data. 
Mc Caulley$^{19}$ has claimed  that 
visible
matter provides no evidence to support either the 
standard cosmological principle
or that the Universe is a fractal/multifractal. 
That may well be true, because 
neither the Cosmological nor the Conditional 
Cosmological Principle are required by
any other known law of Physics. Nevertheless, the 
Cosmological Principle has played
an important indispensible role in development of 
cosmology so far. The Conditional 
Cosmological Principle may play a similar role in the 
fractal scenario. After all, the
idealized homogeneous Universe is a special case of the 
idealized homogeneous fractal
Universe.

	It is hoped that this model will lead to more 
realistic models that incorporate
fluctuations, radiation and nucleosynthesis. All these 
issues need to be scrutinized
afresh from a fractal perspective, looking carefully for 
hidden assumptions
 of homogeneity and continuity in the analysis of 
observed data, specially
because the Standard Model is based on several untested 
physical theories
 and parameter fitting.

\vskip 1cm
\section*{Acknowledgements:}   
   
We thank Inter University Centre of Astronomy and 
Astrophysics
(IUCAA) for hospitality and facilities to carry out this 
research.


\bibliography{plain}

\begin{figure}
   \begin{center}
      \epsfig{file=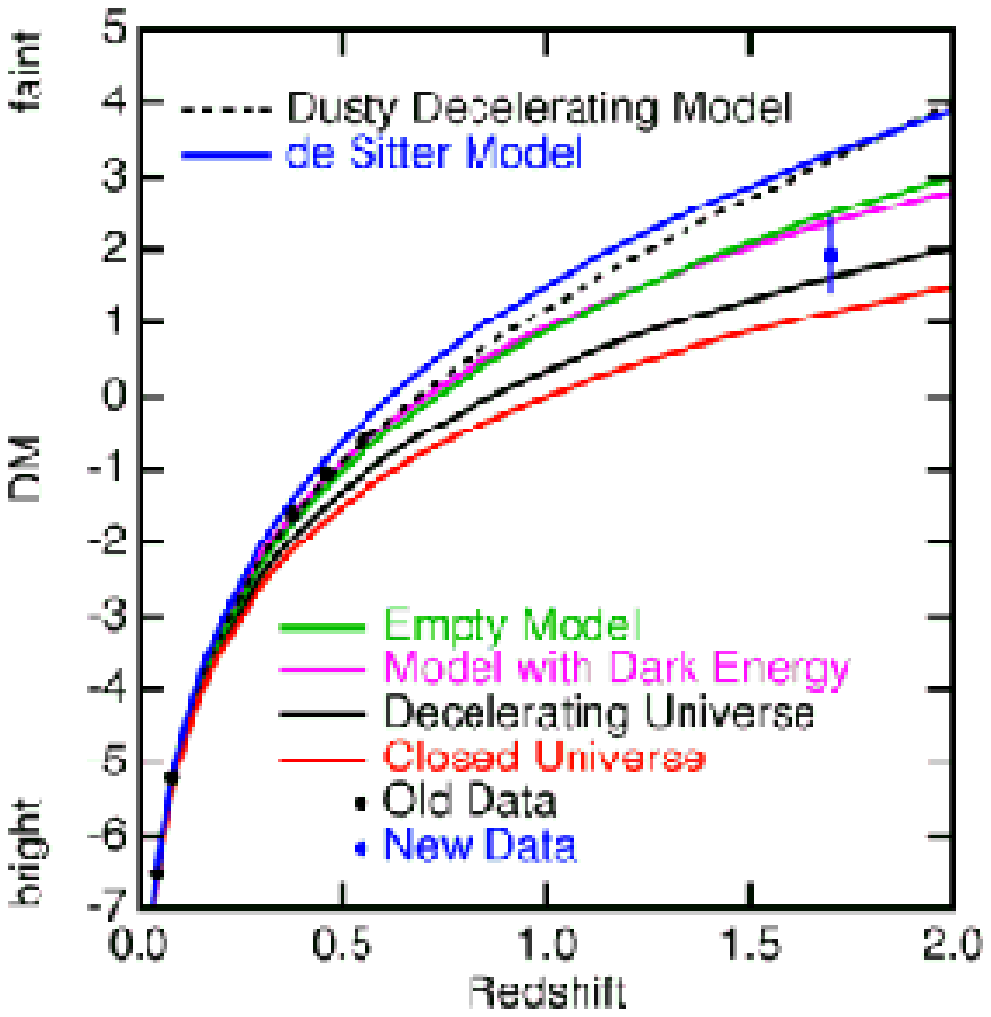, height=10.0cm, width=12.0cm}
   \end{center}
   \caption{Hubble diagram for SNe1a$^{18}$}
   \label{fig1}
\end{figure}

\end{document}